\documentstyle[epsf]{mn}

\newif\ifAMStwofonts

\def\simlt{\lower.5ex\hbox{$\; \buildrel < \over \sim \;$}}
\def\simgt{\lower.5ex\hbox{$\; \buildrel > \over \sim \;$}}

\title[Impact of relativistic corrections and component separation]
    {The impact of relativistic corrections and component separation in 
     the measurement of the SZ effect and on the small angular scale 
     non-Gaussianity of the CMB
     }

\author[Diego, Hansen \& Silk]  {J. M. Diego, S. H. Hansen \& J. Silk\\
    University of Oxford. 
    Denys Wilkinson Building, 1 Keble Road, Oxford OX1 3RH, 
    United Kingdom} \date{Draft version \today}

\begin{document}

\maketitle


\begin{abstract}
We study the effect of imperfect subtraction of the Sunyaev-Zel'dovich
effect (SZE) using a robust and non-parametric method to estimate the
SZE residual in the Planck channels.  We include relativistic
corrections to the SZE, and present a simple fitting formula for the
SZE temperature dependence for the Planck channels.  
We show how the relativistic
corrections constitute a serious problem for the estimation of the
kinematic SZE component from Planck data, since the key channel to
estimate the kinematic component of the SZE, at 217 GHz, will be
contaminated by a non-negligible thermal SZE component. 
The imperfect subtraction of the SZE will have an effect on both 
the Planck cluster catalogue and the recovered CMB map. 
In the cluster catalogue, the relativistic corrections are not a 
major worry for the estimation of the total cluster flux of the 
thermal SZE component, however, they must be included in the SZE 
simulation when calculating the selection function and completeness level. 
The power spectrum of the residual at 353 GHz, where the intensity 
of the thermal SZE is maximum, does not contribute significantly to 
the power spectrum of the CMB.  
We calculate the non-Gaussian signal due to the SZE
residual in the 353 GHz CMB map using a simple Gaussianity estimator,
and this estimator detects a 4.25$\sigma$ non-Gaussian signal at small
scales, which could be mistaken for a primordial non-Gaussian
signature. The other channels do not show any significant departure
from Gaussianity with our estimator.

\end{abstract}


\section{Introduction}\label{section_intro}
The quality of the data from the Planck satellite will allow us to
make precision cosmology, however, accurate parameter extraction will
require precise modeling of the data.  The scientific possibilities
with the new data will depend strongly on the ability to perform the
component separation. In each of the Planck channels the data will be
a mixture of galactic components (synchrotron, free-free and dust),
extra-galactic components (unresolved galaxies and galaxy clusters),
and instrumental noise.  Recently several algorithms have been
proposed to perform such component separation. Most of these methods
rely on {\it a priori} knowledge of the frequency dependence of each
individual component and their power spectrum (maximum entropy,
\cite{Hobson1998}; multi-frequency Wiener filter,
\cite{Tegmark1996,Bouchet1999}).  The advantage of these methods is
that they can recover all the different components simultaneously,
however, the drawback is that if some of the assumptions about the
frequency dependence and/or power spectrum is wrong, then the final
result will be biased.  Other methods are designed to recover just one
of the components, and the most popular ones focus on the recovery of
compact sources. Since the best resolution of Planck is 5 arcmin, the
extra-galactic galaxies will appear as unresolved point sources with a
shape matching the point spread function of the instrument. This fact
can be used to define optimal filters which will increase the signal
to noise ratio of the bright point sources, thus allowing the detection and
removal of most of them \cite{Tegmark1998,Sanz2001}. A similar
technique can be applied to the detection of clusters if one assumes a
circular shape. Since the typical diameter of a cluster is a few
arcmin, most of the clusters will appear similar to the unresolved
point sources, and only some will be resolved by the instrument.  The
definition of the optimal filter is a bit more complicated in this
case since the optimal scale of the filter will be different for each
cluster, however, this problem can be partially solved by filtering
the maps with different scales \cite{Herranz2001}.

An alternative non-parametric approach to detect clusters in Planck
data was recently proposed \cite{Diego2002}.  In that method no
assumptions are made about the specific frequency dependence of the
different components (except for the SZE component), and also no
assumption about the power spectrum of the components or scale (or
symmetry) of the clusters. The only assumption is that the frequency
dependence of the SZE (in the non-relativistic approximation) is
known.  Even without many of the typical assumptions, the recovered
SZE component is a good estimate of the total contribution of galaxy
clusters to the different Planck channels, however, the SZE component
recovered by this method (as well as by other methods) does not
match the real SZE component perfectly in the simulations. Therefore,
a residual will be left in the final CMB map due to the non-perfect
subtraction of the SZE from the original data.  So far there have
been no attempts to study how this SZE residual could affect the
conclusions derived from the {\it residual contaminated} CMB map.  One
of the reasons is that the SZE residual depends on the method used to
make the component separation, and different methods recover the SZE
component with different {\it quality factors} and
residuals. Consequently, the residual SZE map (defined as the ``true''
minus the ``recovered'') is different for each method. In this work we
will study the SZE residual from the non-parametric method proposed in
\cite{Diego2002}.  Since this method makes a minimum number of
assumptions, the results are robust and less affected by systematic
errors which could be introduced through wrong assumptions.  The
conclusions of this work will therefore set an upper limit on the
contributions of the SZE residual to the CMB. Any other method which
makes more assumptions than our non-parametric method, should leave a
smaller SZE residual (provided the assumptions are correct), and as a
consequence the effect of the SZE residual for any other (suitable)
method should be below the limits we present here. It is worth
pointing out that since the non-parametric method adopted here is
optimized for the detection of the SZE signal, then alternative
methods making many more assumptions do not necessarily obtain a
better reconstruction of the SZE signal. Therefore, although our
approach provides an upper limit on the effects of the SZE residual,
then that limit can be taken as a realistic estimate of the final
contribution of the SZE residual to the CMB map.

In this work we will consider an additional source of systematic error
which has not been considered previously, namely the relativistic
corrections to the SZE.  Although these relativistic corrections are
small for normal clusters ($ T \approx $ few keV), they can be
important for massive clusters ($ T \approx 10$ keV). The cluster
selection function of Planck (minimum mass detected as a function of
redshift) rises very quickly from redshift 0 to redshift $\approx$
0.2, and above that redshift it is almost flat (see
Fig. \ref{fig_Selection_Function} below).  This means that above
redshift $\approx$ 0.2 Planck will only see massive clusters for which
the relativistic corrections are important. In all component
separation methods (including the non-parametric method considered in
this work), the frequency dependence of the SZE component is assumed
to follow the non-relativistic form, eq.~(\ref{freqdep}). The validity
of this approach is unknown, and it is therefore worth while exploring
the effect of the relativistic corrections in the context of the
component separation process.

\begin{figure}
   \begin{center} \epsfxsize=8.cm
   \begin{minipage}{\epsfxsize}\epsffile{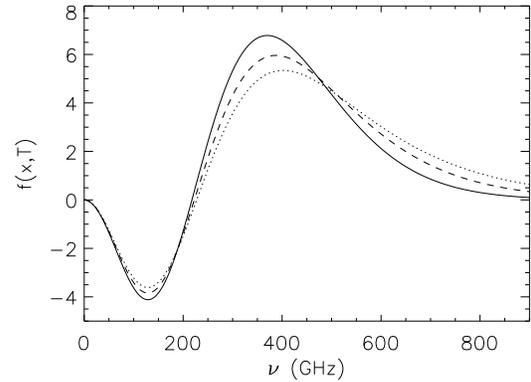}
   \end{minipage}
   \caption{\label{fig_fx_Relat_NoRelat} Solid line is proportional to the
   non-relativistic intensity change which is assumed in all existing
   component separation algorithms. The dashed (dotted) line shows the
   frequency dependence of the SZE when the relativistic corrections
   are included for a cluster with $T = 10$ keV ($T = 20$ keV).}
   \end{center}
\end{figure}

The main effect of the relativistic corrections can be described
effectively as a {\it dilution} of the frequency dependence of the
SZE.  At low frequencies ($\nu < 217$ GHz), the relativistic
correction lowers the absolute value of the SZE intensity change. The
same thing happens at higher frequencies up to $\nu \approx 400$
GHz. Above this frequency, the intensity change is larger than the one
given by the non-relativistic approach (see
Fig. \ref{fig_fx_Relat_NoRelat}), however, the dust contamination is
very important at such large frequencies, and hence the effect of the
relativistic corrections becomes negligible for $\nu> 400$ GHz.

The non-relativistic form is systematically assumed in all component
separation algorithms, however, the dilution effect due to the
inclusion of the relativistic correction makes the form of the SZE for
the hottest clusters different from the non-relativistic form.  This
will lead to an additional error in the residual of the SZE signal.
Our non-parametric method makes the (wrong) assumption that the
relativistic corrections are negligible for all the clusters in our
simulation.  However, we will test the method with SZE simulations
where the relativistic corrections are incorporated. By doing that, we
are simulating the realistic case in which the temperature of the
cluster is not known, and therefore the non-relativistic form must
be assumed in the algorithm to recover the SZE component. This wrong
assumption in our method (and in all other methods) will add an
additional error in the recovered SZE map.

The cluster abundance is a sensitive probe of cosmological parameters
such as the matter density, $\Omega_m$, and when the observations
reach low statistical error it becomes important to control the
systematic errors in the parameter extraction. The relativistic
corrections to the SZE could add an additional error in the estimate
of the cluster number counts, and we will below consider the
importance hereof.

To study the effect of the SZE residual on the CMB we will focus on
two aspects of the CMB, namely its power spectrum and its Gaussian
nature.  The power spectrum depends strongly on the cosmological
parameters, and a systematic error in the estimation of the power
spectrum due to non-subtracted residuals could have important
consequences for the best fitting cosmological model.  Gaussianity is
a natural prediction of single field inflationary models, and a
non-Gaussian signature in the CMB could have important consequences
for such inflationary models.  The SZE signal is very non-Gaussian and
so is the non-subtracted residual.  In previous works (see
\cite{Aghanim1999,Cooray:2001wa,rephaeli2001,Yoshida:2001vf} 
and references therein), the non-Gaussian signature
of the SZE has been studied, but these works focus on the entire
contribution of this component. Since a large part of the SZE signal
will be removed in the component separation process, the non-Gaussian
signature, if any, will be smaller than predicted in previous works,
however, it could still be significant.  It is important to understand
how this residual could leave a non-Gaussian imprint in the CMB map,
such that an erroneous interpretation as of primordial nature of a
non-Gaussian signature can be avoided.  In this work we will study
the implications on Gaussianity studies of the SZE residual left after
component separation.

The structure of the paper is the following. In section
\ref{Section_SZE} we will give a brief description of the SZE and the
relativistic corrections. We also present a simple fitting formula to
the temperature dependence of the SZE in the central frequencies of
the Planck channels.  This fitting formula could be used in future
works to include the relativistic corrections in the simulations.  In
section \ref{section_residual} we apply the non-parametric method to
realistic Planck simulations and we recover the SZE component in two
cases.  In the first case we simulate a population of clusters all
with the same temperature.  This allows us to assume the real
frequency dependence of the SZE in the non-parametric component separation 
method and see what the difference is with the case when the non-relativistic
form is assumed in the component separation.  In the second test,
we make realistic simulations of the clusters with the population of
clusters having different temperatures. We include the relativistic
correction in our simulation. Then we calculate the SZE residual
assuming the non-relativistic approach for the frequency dependence of
the SZE.  In section \ref{section_EffectSZE} we discuss the effects of
the imperfect SZE recovery on the Planck cluster catalogue focusing
our attention on the systematic errors introduced in the recovered SZE
map by the relativistic corrections and the problems this causes for
the kinematic SZ component.  In section \ref{section_EffectCMB} we
discuss the effect of the SZE residual on the CMB map with emphasis on
non-Gaussian signatures.  Finally we present our conclusions in
section \ref{section_Conclusions}.

\section{The Sunyaev-Zel'dovich effect with relativistic corrections}\label{Section_SZE}
In this section we will give a brief description of the SZE
and the relativistic corrections (see e.g. the reviews 
\cite{Birkinshaw99,Carlstrom01} for more details).

As the photons of the cosmic background radiation (CMB) traverses a
cluster of galaxies they may scatter on the free electrons in the
ionized gas and produce the thermal Sunyaev-Zel'dovich
effect~\cite{sz}.  The resulting intensity change of the CMB is
proportional to the Comptonization parameter,
\begin{equation}
y_c = \int dl~ \frac{T_e}{m_e} ~ n_e \sigma_{\rm Th} ~,
\end{equation}
where $T_e$ is the temperature of the electron gas in the cluster,
$m_e$ the electron mass, $n_e$ the electron number density,
$\sigma_{\rm Th}$ the Thomson scattering cross section, and the
integral is calculated along the line of sight through the cluster. We
use units where $k_B=\hbar=c=1$.  For an intra-cluster gas which can
be assumed isothermal, one has $y_c=\tau \, T_e/m_e$, where $\tau$ is
the optical depth.

The intensity change is given by
\begin{eqnarray}
\Delta I_{\rm T} &=& 
I_0~ y_c~ f(x,T)~,
\label{thermal}
\end{eqnarray}
with
\begin{equation}
f(x,T) = f(x) + \delta f(x,T_e)
\label{deltaf}
\end{equation}
and
\begin{equation}
f(x) = f(x,0) = \frac{x^4e^x}{(e^x-1)^2}\left[
\frac{x(e^x+1)}{e^x-1}-4\right] \, ,
\label{freqdep}
\end{equation}
where $x=\nu/T_{\rm CMB}$ is the dimensionless frequency ($T_{\rm CMB}
=2.725 ~{\rm K}$), and $I_0=T_{\rm CMB}^3/(2\pi^2)$. The intensity
change is independent of the temperature for non-relativistic
electrons, $\delta f(x,T_e)=0$, a limit which is valid for small
frequencies ($\nu\simlt 100$ GHz), but for high frequencies it must be
corrected ~\cite{wright1979,rephaeli95} with $\delta f(x,T_e)$ using either an
expansion in $T_e/m_e$ \cite{stebbins97,Challinor98,ItohI,ItohV} or
calculated exactly \cite{dolgov00}. One can use these relativistic
corrections to find the temperature of distant
clusters~\cite{hansen02}, and possibly even to find the temperature of
clusters in the Planck catalogue~\cite{pgb98}.

Since the exact calculation of the relativistic correction is time
consuming, it is more convenient to use a fit to this correction as a
function of the temperature. We have calculated such a fit in the
range $T \in [0.5,20]$ keV and we found that a fit of the form,
\begin{equation}
f(x,T) = \alpha_\nu + \beta_\nu \, T +
\gamma_\nu \, T^2
\end{equation}
is rather accurate (where $T$ is given in keV).  Naturally one should
have $\alpha_\nu = f(x,0)$, however, since the fit is optimized for
$T=1-20$ keV, the $T\approx 0$ limit is slightly different.  We use
the exact calculations of Dolgov et al. (2001) and fit for each of the
central Planck frequencies. The resulting parameters $\alpha, \beta$
and $\gamma$ are presented in table 1~\footnote{This table (and
practical details) can be found on {\tt
http://www-astro.physics.ox.ac.uk/\~{}hansen/sz/}}.

\begin{table}
\begin{tabular}{c c c c} 
\hline
$\nu$ (GHz)   &  $\alpha_\nu$ & $\beta_\nu$        & $\gamma_\nu$ \\ 
\hline

30   &  -0.531297   &  0.00186392    & -2.04638E-06\\
44   &  -1.08285    &  0.00423301    & -5.5792E-06\\
70   &  -2.3448     &  0.0116543    &  -3.38278E-05\\
100   & -3.62693    &  0.0227427    &  -0.000115648\\
143   & -3.98417    &  0.0280639    &  -0.000230855\\
217   & -0.0527698  &  -0.0338395   &   0.000389163\\
353   &  6.68822    &  -0.105281    &   0.000987777\\
545   &  3.25014    &  0.0585885    &  -0.0015262\\
857   &  0.152158   &  0.0302537    &   0.000126894\\

\hline
\end{tabular}
\label{table_1}

\vspace{\baselineskip} {\small T{\scriptsize ABLE}~1 --- Fit
parameters $\alpha_\nu, \beta_\nu$ and $\gamma_\nu$ for the expected
Planck central frequencies. For each frequency the coefficients come
from $\Delta I_{\rm T}(\nu)/(I_0~ y_c) = \alpha_\nu + \beta_\nu \, T +
\gamma_\nu \, T^2$ where $T$ is measured in keV.}
\end{table}

It is important to keep in mind, that these fit parameters have been
calculated while neglecting bandwidth. In a real observational
situation the proper inclusion of bandwidth is crucial, e.g. a
Gaussian frequency bandwidth of 35 GHz will produce $5\%$ less
intensity change for the 353 GHz channel. However, this intensity
change is about $5\%$ {\em both}, when considering only the
non-relativistic form and when considering the full relativistic
treatment, so the results presented in this paper would basically be
identical when including a bandwidth.

\section{The recovered SZE map and its residual}\label{section_residual}
Since our present knowledge of
the galactic and extra-galactic components is limited, the proposed
algorithms for component separation will leave a galactic and
extra-galactic residual in the CMB map, and as a consequence the power
spectrum of the CMB will be distorted.  This distortion will be
particularly important at small scales ($\theta \approx 5$ arcmin, $l
\approx 2000$).  The physics extracted from this part of the power
spectrum will therefore be limited by the accuracy achieved in the
component separation or {\it cleaning} process which will add a
systematic error in the CMB power spectrum.

In this work we want to investigate one of these sources of systematic
errors, namely the imperfect recovery of the SZE including
relativistic corrections.  Usually it has been assumed in component
separation algorithms that the SZE frequency dependence is described
by the non-relativistic form, eq.~(\ref{freqdep}). There are several
reasons to make this assumption. First, the relativistic correction is
small for many clusters.  Second, the temperature of the clusters is
unknown for almost all the clusters and consequently it is not
possible to calculate the relativistic correction for the individual
cluster.  Finally, the different component separation algorithms are
much simpler if the same frequency dependence is assumed for all the
clusters in the map.  However, the non-relativistic assumption will
only be valid for those clusters where the temperature is small
(typically a few keV).  For massive (hot) clusters the relativistic
correction can be important.  For instance, for a cluster with $T
\approx 10$ keV , this correction is about 15 \% at frequencies
near 353 GHz (see table 1), and consequently there will be an
additional error in the estimate of the signal of the hot
clusters. This error will leave an imprint in the CMB when the
incorrectly estimated signal of these hot clusters is removed.  Since
most of the clusters will be unresolved, this error will contribute
only to the small scales of the CMB. Whether or not this error can
distort the power spectrum of the CMB significantly at small scales is
still an open question.

The effects of the non-relativistic assumption could also be important
for the studies based on the cluster number counts. Since the
non-relativistic assumption will introduce some error in the estimate
of the flux of hot clusters, the number counts as a function of flux
will be affected by this error. It is important to quantify this error
when such data sets are going to be used to extract cosmological
parameters.

It is therefore important to investigate with simulations whether or
not the effects of the non-relativistic assumption are important in
the component separation process and if they could represent a serious
problems for the estimate of the cosmological model which best
describes the CMB data.
The simulations of the Planck data used in this work are the same as
those described in \cite{Diego2002} and the reader is
referred to that paper for more details. 
Our simulations include the following components;
galactic emission (synchrotron, free-free, and dust), point sources,
SZE, CMB and instrumental noise. We also take into account the
different resolutions in each channel. The simulations correspond 
to an area of the sky of $12.8^{\circ} \times 12.8^{\circ}$, and each 
map contains  $512^2$ pixels.
The cosmological model assumed for the CMB and the SZE simulation is 
consistent with recent CMB and cluster abundance observations 
($\Omega_m = 0.3$, $\Lambda = 0.7$, $n=1$, $\Gamma = 0.2$, 
$\sigma_8 = 0.8$).

\subsection{Component separation. Non-parametric method}
We will here give a brief summary of how to subtract the SZE component
from the simulated Planck maps.  A detailed description of the method
can be found in \cite{Diego2002}. 

The method is basically a non-parametric Bayesian approach.  It is
non-parametric because it does not make any of the typical assumptions
about the components, such as knowledge of the power spectrum and
frequency dependence. The only assumption made in this method is that
we know the frequency dependence of the SZE. It also makes the (wrong)
assumption that the SZE signal is a Gaussian variable. However, as
described in \cite{Diego2002}, when this assumption is made over
the Fourier modes of the Compton parameter map, these modes follow a
probability distribution function which is much closer to a Gaussian
than in real space. The method performs a basic cleaning of the maps
where a first estimate of the point source, dust and CMB contributions
is removed. Then the Bayes theorem is applied on the {\it clean} maps
and an estimate of the SZE map is obtained.
\begin{equation}
y_c = \frac{{\bmath d}{\bf C}^{-1}{\bmath R}^{\dag}}{{\bmath R}{\bf C}^{-1}{\bmath R}^{\dag} + P_{y_c}^{-1}}.
\label{eq_Search_engine}
\end{equation}
where $y_c$'s are the Fourier coefficients of the Compton parameter
map.  ${\bmath d}$ is the vector containing the Fourier coefficients of the
{\it clean} maps, ${\bf C}^{-1}$ is the inverse of the correlation matrix of
the {\it clean} maps, ${\bmath R}$ is another vector (same dimension as
${\bmath d}$) containing the Fourier transform of the beam response of
the instrument and the frequency dependence of the SZE, and $P_{y_c}$
is an estimate of the power spectrum of the Compton parameter
map. This estimate can be obtained from the data after a first run of
the code.  Finally, the Compton parameter map is obtained by Fourier
transforming the Fourier coefficients obtained in 
equation~(\ref{eq_Search_engine}).

\subsection{The case of known temperature}
In the Bayesian approach, one often assumes knowing the frequency
dependence of the SZE, and that this frequency dependence follows the
non-relativistic form, i.e $\delta f(x,T_e)=0$ in
eq.~(\ref{deltaf}).  Since the frequency dependence of the
relativistic correction depends on the unknown temperature of the
cluster this cannot be computed when the temperature of the cluster is
unknown. This assumption is common to all the existing
component separation algorithms.
\begin{figure*}
   \begin{center} \epsfxsize=18.cm
   \caption{\label{fig_SZ_Recov_True_353GHz_Diff} Recovered SZE map
   (bottom left) compared with the input SZE map used in the
   simulation (top left) in the 353 GHz channel. 
   The right large panel shows the difference
   between both maps (the residual). The units are $\Delta T/T$ at 353
   GHz. The brightest point in the residual is 
   about 50 \% of the original value.  
   These bright points are located in the hottest clusters. 
   This residual will mainly contaminate the CMB
   component. } \end{center}
\end{figure*}
To test how important is the error introduced in the recovered SZE map
due to the previous assumption, we have simulated a population of
hot clusters at different redshifts (up to $z = 1$) and with different
masses, but with the constraint that all of them must have the same
temperature, $T = 10$ keV.  In our simple case, where we know the temperature 
$T=10 KeV$, we can use the real (including the relativistic corrections) 
frequency dependence of the SZE in the component separation algorithm since 
the SZE map is composed of clusters all with the same T and consequently, 
all with the same $f(x,T)$.  Then, we apply our Bayesian estimator to our
Planck simulations which includes galactic components, CMB, point
sources, SZE and instrumental noise, and we recover the SZE component
in two situations. In the first case we assume the real $f(x,T)$
(including the relativistic correction) and in the second case we
assume that $f(x)$ corresponds to the non-relativistic case. Then we
look at the difference of the two recovered maps. We find that the
relative difference in the two Compton parameter maps is between
$\approx 4\%$ (low redshift clusters, $z<0.5$) and $\approx 8\%$ (high
redshift clusters, $z>0.5$).  This difference is smaller than the 15\%
difference between the relativistic and non-relativistic approaches at
353 GHz ($T = 10$ keV). The reason for that is that the recovered SZE
map is obtained by {\it averaging} over different channels where the
differences between the relativistic and non-relativistic forms
are smaller than 10\% for the channels below 353 GHz (the two channels at 545 and 857 GHz
do not contribute significantly to the previous
average). We can conclude that the effect of the non-relativistic
assumption in the component separation algorithm is not a major worry
(but see section \ref{section_EffectSZE} below). Therefore, in the
case of real life where the population of clusters have different
(unknown) temperatures, the non-relativistic form can be assumed
in the component separation since it will only introduce a small
error.  As we will see in the next section, the main error in the SZE
estimation will come from the imperfect separation between the SZE and
the other components.

\subsection{The real case. Unknown temperatures.}\label{section_Real}
In the previous section we have seen that for a population of hot
clusters ($T = 10$ keV) for which the relativistic corrections can be
important, the effect of assuming the non-relativistic form in the
component separation algorithm is not larger than 8\%. However, even
in the case when we assume the right frequency dependence, the
component separation algorithm will have an intrinsic error and the
SZE will not be subtracted perfectly.  This case was considered in
\cite{Diego2002} but in that work the authors did not include the
relativistic corrections neither in the simulation of the SZE
component nor in the assumed frequency dependence of the SZE in the
non-parametric method.

When testing component separation algorithms with simulations, the
frequency dependence of the SZE appears not only in the assumption made
in the component separation process but also in the simulation of the
SZE component.  Usually the simulations of the SZE do not include the
relativistic correction, and as explained in the introduction, the
{\it dilution} effect implies that if the simulations do not include
the relativistic correction, clusters will appear with a greater contrast at
the most relevant frequencies for cluster detection. In real life the
clusters will be dimmer due to this {\it dilution} effect.  In this
section we will include the relativistic correction of the SZE in our
simulation as described in table~1, and we will compare with the case
where the correction is not considered in the simulation. The cluster
distribution in the $M-z$ space is obtained from
Press-Schechter~\cite{ps} for a flat $\Lambda$CDM universe with
$\Omega_m = 0.3$ and $\Omega_{\Lambda} = 0.7$. The temperature of the
clusters is obtained from the $T-M$ relation. For this relation we have used 
the fitting formula found by \cite{Diego2001} where the authors fitted 
the $T-M$ relation to several X-ray data. 
The electron density is assumed to follow a standard 
$\beta -$model ($\beta = 2/3$), and we
use $\sigma_8=0.8$ (for more details see \cite{Diego2002}). Once we
have the temperature of the cluster, the relativistic correction is
computed for each cluster and in each one of the Planck channels.  In
the simulated area of the sky ($12.8^{\circ} \times 12.8^{\circ}$),
there are about 20000 clusters in the simulation above a mass of $3
\times 10^{13} h ^{-1} M_{\odot}$ which corresponds to a temperature
$\approx$ 1 keV.  From those clusters only a few tens will be
detected by Planck in this area of the sky. The hottest cluster in
this simulation has a temperature $T = 15$ keV. The most massive
cluster in the simulation has $M = 1.1 \times 10^{15} h^{-1}
M_{\odot}$, and it is at $z = 0.66$.

In Figure~\ref{fig_SZ_Recov_True_353GHz_Diff} we present this
simulated SZE map compared with the SZE map recovered by the
non-parametric method (assuming the non-relativistic form for $f(x)$
in the component separation process) and the difference between them
(the residual) in the 353 GHz channel.  We do not present here the
case where the relativistic correction is not used in the simulation
because it looks very similar to 
Figure~\ref{fig_SZ_Recov_True_353GHz_Diff} (see \cite{Diego2002}).  The
recovered map contains positive and negative values. This is due to
the {\it wrong} choice of SZE prior in the non-parametric method.
When computing the residual (true SZE map minus recovered SZE map),
the negative values in the recovered map must be set to 0 since the
Compton parameter must contain only positive values.

In the next section we will compare the recovered SZE map
(Figure~\ref{fig_SZ_Recov_True_353GHz_Diff} bottom left) in the two
cases, SZE simulation including the relativistic correction and SZE
simulation not including the relativistic correction.  In section
\ref{section_EffectCMB} we will concentrate on the residual
(Figure~\ref{fig_SZ_Recov_True_353GHz_Diff} right large panel) in the
case with relativistic correction and consider the effect on the CMB.

\section{Planck cluster catalogue}\label{section_EffectSZE}
\subsection{Thermal SZ}
The recovered map (Figure~\ref{fig_SZ_Recov_True_353GHz_Diff}
bottom-left) is a {\it noisy} estimate of the real map
(top-left). This noisy map is composed basically of two components. A
Gaussian background which contains most of the spurious signal in the
recovered map and a positive tail (in the pdf) which contains the
galaxy clusters. In order to discriminate between the noisy background
and the clusters we have used the package \small{SEXTRACTOR}
\normalsize \cite{Bertin1996}. 
With \small{SEXTRACTOR} \normalsize we can detect 46
clusters (above a threshold of 3$\sigma$ and with 10 pixels
connected). We have compared this number of detections with the ones
obtained when the original SZE map is simulated without considering
the relativistic correction. In this case, the assumption made in the
non-parametric method about the non-contribution of the corrections to
the SZE frequency dependence is correct and there is no systematic
error introduced by the relativistic corrections.

\begin{figure}
   \begin{center} \epsfxsize=8.cm
   \begin{minipage}{\epsfxsize}\epsffile{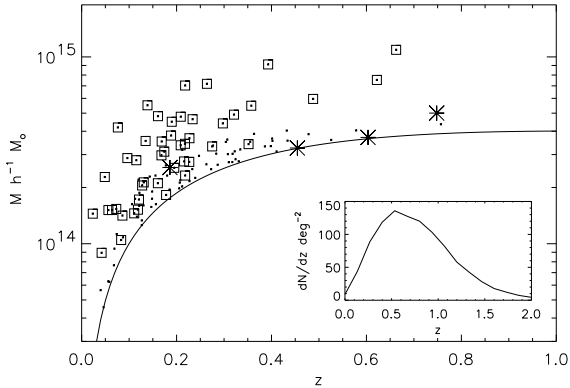}\end{minipage}
   \caption{\label{fig_Selection_Function} Each dot in this plot
   represents a cluster in our simulation (only clusters with fluxes
   bigger than 60 mJy are plotted). The solid line represent the
   selection function at 60 mJy. Dots surrounded by a square 
   are the clusters detected using the Bayesian approach. The 4 
   clusters marked with a big star are the clusters which are not
   detected by {\small{SEXTRACTOR}} when the relativistic corrections
   are included in the SZE simulations. The insert plot shows the underlying 
   distribution of clusters for this model ($dN/dz$). } 
   \end{center}
\end{figure}

When the simulated SZE does not include relativistic corrections the
number of detections returned by \small{SEXTRACTOR} \normalsize (with
the same criteria as above) is 50. In
Figure~\ref{fig_Selection_Function} we show the recovered clusters in
both cases (SZ simulation with and without relativistic
corrections). We also show the selection functions at the flux 
60 mJy. Above 200 mJy (at 353 GHz) the method detects almost 100 \% of
the clusters.  No clusters are detected below the flux 60 mJy.  There
are 4 {\it missing} clusters which are not detected when the SZE
simulation includes the relativistic corrections (big stars).  This is
just an example of the {\it dilution} effect of the frequency
dependence when the relativistic corrections are included. The hot
clusters become less {\it bright} in the main Planck channels when the
relativistic correction is included and it is therefore more difficult
to detect them.  It is important to note that these missing clusters
are more concentrated around intermediate-high redshifts 
($z \approx 0.2-0.8$).  
The {\it dilution} effect is more important for the hottest clusters which, 
for the same mass, are expected to be in the high redshift interval 
($T \propto M^{2/3}(1+z)$ in the $\Omega_m = 1$ case (note that in the 
simulations we use the more general formulae with $\Omega_m=0.3$ and 
$\Omega_\Lambda=0.7$)).


The fact that the missing clusters are close to the limiting flux of the 
survey makes the relativistic correction an important issue to be taken 
into account in future modeling of the data.  
The future catalogue of clusters obtained by Planck could be
used as an independent cosmological test, however, in order to do that
it is crucial to understand the selection function and completeness
level of the catalogue. Our results show that an accurate estimation
of both selection function and completeness obtained from simulations
should include the relativistic corrections in the simulations in
order to not underestimate these quantities at intermediate/high redshifts.

\begin{figure}
   \begin{center} \epsfxsize=8.cm
   \begin{minipage}{\epsfxsize}\epsffile{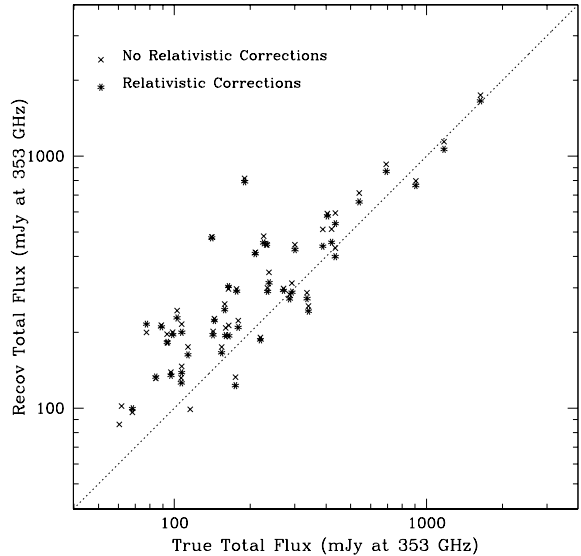}\end{minipage}
   \caption{\label{fig_Flux_True_vs_Recov_Both} Recovered fluxes
   versus true fluxes at 353 GHz. Crosses are the fluxes when the
   relativistic corrections are taken into account in the simulated
   SZE map (but not in the Bayesian approach). Asterisks indicate the
   recovered fluxes when both in the simulation and in the Bayesian
   approach, the relativistic corrections are not included. 
   } \end{center}
\end{figure}

The relativistic corrections will not only have an effect on the
number of clusters which can be detected but also in the accuracy in
the estimation of their fluxes.  In 
Figure~\ref{fig_Flux_True_vs_Recov_Both} we show how well \small{SEXTRACTOR}
\normalsize recovers the total flux of the cluster in both cases (with
and without the relativistic corrections in the simulated SZE map).
The difference between considering and not considering the
relativistic correction in the SZE simulated map is small in the computation of 
the total flux.

\subsection{Kinematic SZ}
\begin{figure}
   \begin{center} \epsfxsize=8.cm
   \begin{minipage}{\epsfxsize}\epsffile{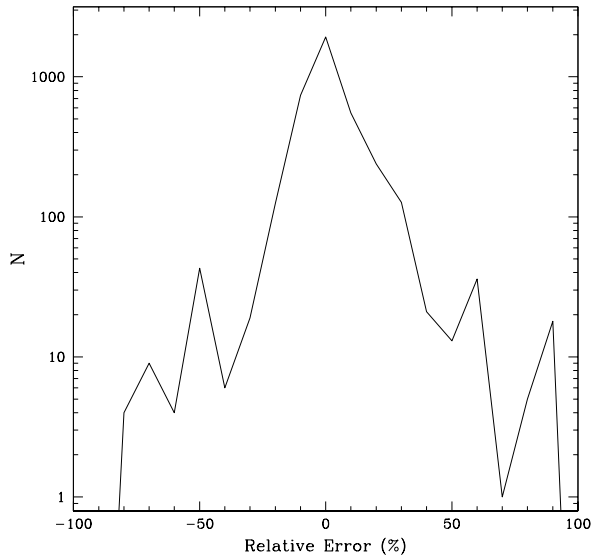}\end{minipage}
   \caption{\label{fig_RelatError_Kin} The histogram shows the number
   of pixels above threshold (see text) as a function of their
   relative error in the kinematic component due to the non-subtracted
   thermal component. We have chosen a broad binning (20 bins) in order to 
   have a significant number of points in the tails of the distribution.}  \end{center}
\end{figure}

The previous discussion was for the thermal component of the SZE, but
the kinematic component will also be affected. In this case the
distortion in the CMB temperature is given by
\begin{equation}
\frac{\Delta T}{T}_{Kin} = - \frac{v_r}{c} \tau
\end{equation}
where $v_r$ is the radial (peculiar) velocity, and $\tau$ the optical
depth of the cluster.  The kinematic SZE component has the same
frequency dependence as the CMB fluctuations. Furthermore, its intensity is
typically 20-30 times smaller than the thermal SZE component.  This
makes it extremely difficult to measure the kinematic component. One
of the best strategies to detect it is to apply an optimal filter for
cluster detection (see Herranz et al. 2002) to the 217 GHz channel
where the thermal SZE component vanishes (in the non-relativistic
approach) and then cross-correlate the filtered map with the positions
where the thermal SZE component was detected.  However, when we
consider the relativistic corrections, we cannot simply assume that the
thermal SZE component vanishes at 217 GHz (the
cross-over~\footnote{The cross-over frequency goes like $f_0 = 217.5(1
+ 2.27 \times 10^{-3} \, T - 3.12 \times 10^{-6} \, T^2) $ for
vanishing optical depth, $\tau\approx0$.} is at $\approx 225$ GHz for
T = 15 keV).  Instead, there will be a contribution of the thermal SZE
at this frequency which can be of the same order of magnitude of the
kinematic SZE.  In order to quantify how important the contamination
of the thermal SZE to the kinematic component is in the 217 GHz
channel, we have compared a simulated SZE map of the kinematic effect
with the corresponding thermal component (including relativistic
corrections) at this frequency.  The kinematic component has
been simulated assuming a Gaussian distribution for the radial
velocities with a dispersion of $\sigma _v = F(z)*400$ km/s, where
$F(z)$ accounts for the evolution of the velocity field with redshift
in linear perturbation theory, (Peebles 1980)
\begin{equation}
F(z) = \frac{\dot{D}(z)}{\dot{D}(0)} \frac{H}{Ho}(1 + z)^{-2} \, .
\end{equation}
The rest of the parameters are the same as in the simulation of the
thermal component.  In Fig. \ref{fig_RelatError_Kin} we present the
relative error which, in each pixel, has been defined as
\begin{equation}
Err = 100 \times \frac{\Delta _T}{\Delta _K} \, ,
\end{equation}
where $\Delta _T$ is $\Delta T/T$ for the thermal component and
$\Delta _K$ for the kinetic one. This relative error is $\approx$ 0 in the
non-relativistic approach.  Since the radial velocity of the clusters
can be arbitrarily small, we have to set a threshold in the
calculation, which is $\Delta T/T = 10^{-6}$ for the
kinematic effect, and we only compute the relative error for clusters
above this threshold.
As can be seen in figure 5, the relative error can be as large as
$100\%$ for some clusters and there are several clusters with relative
errors $\sim 50 \%$. This will constitute a serious problem for the
correct estimation of the kinematic SZE component in those clusters,
and thus making it harder to constrain theories of structure formation
and evolution from measurements of the radial peculiar velocities (see
e.g. \cite{Aghanim2001}).  Since the relative error is proportional to
the thermal SZE component which includes the relativistic corrections,
and since the relativistic corrections are more important for the
hottest clusters, the relative error is (in general) larger for the
hottest clusters and smaller for the coldest ones.

In our calculations, we did not consider the bandwidth for Planck in
the 217 GHz channel. However, our conclusions are correct if the
bandwidth is symmetric around the central frequency since in this case
the effect of integrating along the bandwidth is the equivalent to
taking the central frequency, because the frequency dependence
of the SZE can be well approached by a straight line over the
bandwidth at 217 GHz.

\section{The CMB map}\label{section_EffectCMB}
The SZE residual in Figure~\ref{fig_SZ_Recov_True_353GHz_Diff} will
contaminate the other components, and in particular it will
contaminate the most important component (from a cosmological point of
view), the CMB.  The percentage of the SZE residual which contaminates
the CMB will depend on the method used to perform the component
separation.  There are two extreme cases which can be considered. One
in which no SZE residual contributes to the CMB map and the other case in
which 100 \% of the SZE residual contaminates the CMB.  The real
situation will be between the two extreme cases, and only part of the
SZE residual will contaminate the CMB map.  \\

There is one basic constraint in all the component separation algorithms which 
must be obeyed. The sum of all the components (in each pixel) must be equal to
the data.  Since most of the component separation algorithms work in
Fourier space, this constraint must be reformulated in terms of the
Fourier coefficients.  The SZE residual is basically contributing at
small scales. This means, that at these scales we have not subtracted
all the SZE signal. Due to the previous constraint, this residual will
go to some of the other components and will contaminate the Fourier
modes at the small scales. To understand which components are most
affected by the SZE residual, we have to understand the way the
different component separation methods work.  In typical 
component separation methods (and in our method too), the point
sources are first subtracted using an optimal filter (or an equivalent
one). Since the SZE has not been removed yet, no SZE residual can go
to the point source component. Then the rest of the components are
separated assuming some frequency dependence for all of them and a
correlation matrix between the components (in our method we do not
need to make these assumptions). This cannot be done in the case of
point sources since they do not have the same frequency dependence.
This is the reason why they have to be subtracted first.  It is
usually assumed in the correlation matrix that the galactic components
(synchrotron, free-free and dust) have much more power at large scales
than at small scales. In our method we do not make this
assumption. Instead the dust is partially removed by subtracting the
857 GHz channel times a constant which minimizes the variance of the
difference, however, the situation is similar to that described above,
i.e, the dust has much more power at large scales than at the small
ones and when we remove the dust, we are basically removing a diffuse
component at large scales while the small scales do not change
substantially.  This means that the power at small scales is basically
due (after point source subtraction), to the contributions of the SZE
and the CMB (at even smaller scales the instrumental noise dominates
over the rest of the components).  We have thus seen that the major
contaminant of the CMB at small scales is the SZE. If we do not
remove perfectly the SZE component at small scales, its residual will
contaminate the CMB modes at those scales.
 
In the previous discussion, we have assumed that the removal of the
point sources is perfect. In reality the point source subtraction
produces another residual which will contribute to small scales. This
residual will have an effect on both the CMB and the SZE, however, the
effect on the SZE component is small since point sources have a quite
different frequency dependence from the SZE. Therefore, the point
source residual will basically affect the CMB at small
scales. However, this point is beyond the scope of this paper and we
will concentrate only on the SZE residual.

\begin{figure}
   \begin{center} \epsfxsize=8.cm
   \begin{minipage}{\epsfxsize}\epsffile{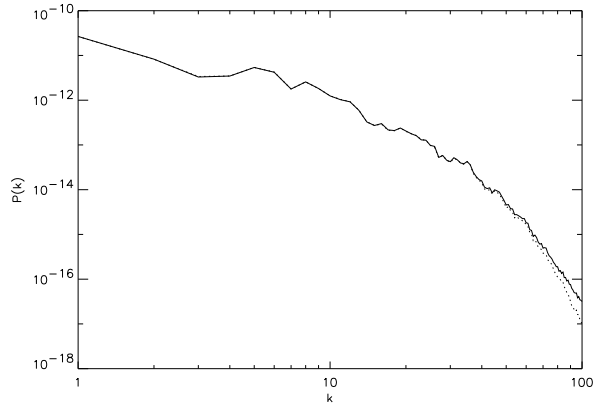}\end{minipage}
   \caption{\label{fig_Power_TrueCMB_Sum} Power spectrum of the CMB
   map (dotted line) and power spectrum of the sum CMB plus SZE
   residual (solid line). The effect of the residual is small in the
   power spectrum.  }  \end{center}
\end{figure}

Now the question is how important is the non-subtracted SZE signal
(the residual) for the CMB science. The basic quantity in cosmological
studies based on the CMB is the power spectrum.  In order to check the
relevance of this residual in the CMB we have computed that quantity
in the two extreme cases. Where the CMB map does not have any residual
and when the CMB map includes all the SZE residual.  We also compute
these quantities in the channel at 353 GHz where the SZE residual is
the largest (the channels at 545 and 857 GHz cannot be used to obtain
the CMB power spectrum). By taking these two extreme cases we can set
an upper limit on the effect of the SZE residual.  The result is shown
in Figure~\ref{fig_Power_TrueCMB_Sum}.  As can be seen from that
figure, even in the extreme case where all the SZE residual is
contained in the CMB component, the power spectrum is not very much
affected by the SZE residual. These are good news since one does not
have to worry much about the contribution of the SZE residual to the
power spectrum. Only at very small scales, the distortion in the power
spectrum due to the residual can be important but these high-$k$ modes
will be dominated by the error bars. These error bars include
instrumental noise, cosmic variance, and the error associated with the
component separation process. If one takes into account all the
previous error bars one finds that Planck will be able to estimate the
power spectrum of the CMB up to $k \approx 70$ ($l \approx 2000$). At
those scales, the contribution of the SZE residual to the power
spectrum is almost negligible.
%
\subsection{Non-Gaussianity}
There are, however, other studies which can be carried out with CMB
data apart from its power spectrum.  Gaussianity studies are very
important since Gaussianity of the CMB is a basic prediction of single
field inflationary models.  As an alternative to inflationary models
there are e.g. models in which the structure originates from
topological defects, and such models predict a non-Gaussian pattern
for the CMB.  When mixed adiabatic and isocurvature models are
compared with current CMB power spectrum, one finds that maximum
$15\%$ isocurvature is allowed by the data \cite{Enqvist2000}, 
while pure adiabatic models give a very good fit.
For Gaussianity studies one should be extremely careful when dealing
with CMB maps. Even if the real CMB map is Gaussian, the map will
contain some residuals which can be non-Gaussian. In the case of the
SZE the residual is clearly non-Gaussian. It is, therefore, important
to see if the non-Gaussian signal due to the SZE residual is relevant
or not.

\begin{figure}
   \begin{center} \epsfxsize=8.cm
   \begin{minipage}{\epsfxsize}\epsffile{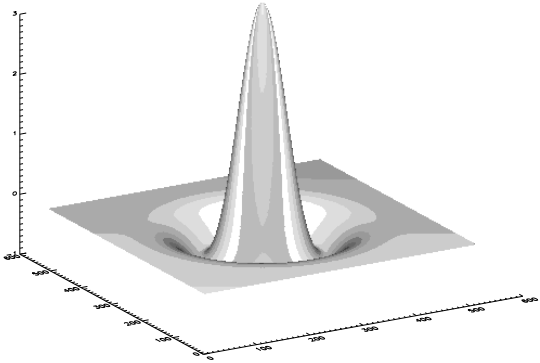}\end{minipage}
   \caption{\label{fig_MHW} 
    MHW in real space. The size of the grid over which the MHW is defined in this plot 
    is $30'\times 30'$.      
    Although the scale, $s$, of this MHW is 3 arcmin, then the total size of the MHW 
    is $\approx 20$ arcmin.  
    } 
   \end{center}
\end{figure}

\begin{figure}
   \begin{center} \epsfxsize=8.cm
   \begin{minipage}{\epsfxsize}\epsffile{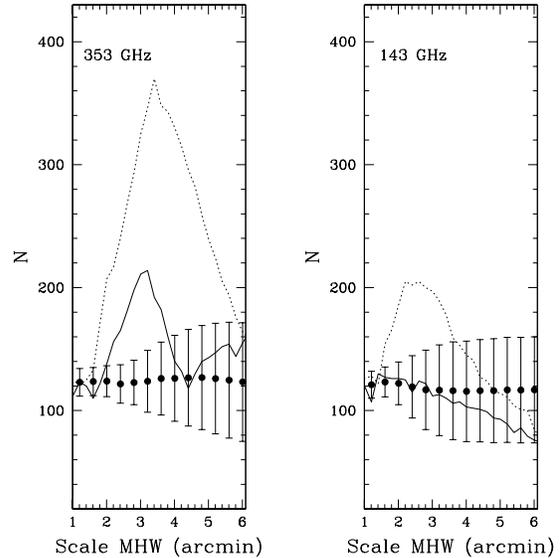}\end{minipage}
   \caption{\label{fig_N_Threshold_MHWF} Number of MHW coefficients
   above $3.5\sigma$ level as a function of the scale of the MHW for two 
   Planck channels, 353 GHz (left) and 143 GHz (right).  The
   solid line is the number of coefficients (in absolute value) above the 
   threshold for the Gaussian CMB map plus the SZE residual plus noise.
   For reference, we also show the case
   when the SZE is not removed at all (dotted line).  The big
   black dots are the mean value of 100 Gaussian realizations of the
   CMB plus the corresponding noise at that channel and the error bars 
   are $1\sigma$. The mean value is
   consistent with the expected number of points above $3.5\sigma$ for
   a map with $512^2$ wavelet coefficients (in absolute value).  
   This plot clearly shows,
   how the residual adds a significant ($> 4\sigma$) non-Gaussian
   signature at small scales in the 353 GHz channel. 
   In the case of the 143 GHz channel, the SZE residual does not introduce a 
   significant non-Gaussian signature. In the rest of
   the channels, the number of coefficients above $3.5\sigma$ are also 
   consistent with the Gaussian case. }  \end{center}
\end{figure}

The power spectrum of the CMB is a good indicator of the {\it mean}
contribution of the signal at different scales. However, the power
spectrum is not an estimator of the Gaussianity of a map, and one has
to use other estimators. If we look at the residual in 
Figure~\ref{fig_SZ_Recov_True_353GHz_Diff} we see that the main contribution
of the residual is localized in compact peaks. If the CMB map contains
this residual, then the peaks of the residual will be mixed with the
intrinsic peaks of the CMB. A good Gaussian estimator in this case
should concentrate on the small regions where the compact peaks of the
residual are more likely to contribute.  We propose the use of the
mexican hat wavelet (MHW) although other filters could be used
instead.  The MHW in real space is given by the second derivative of a
Gaussian;
\begin{equation}
\Phi(r) = (2.0 - (\frac{r}{s})^2) exp(-\frac{r^2}{2 s^2}) 
\end{equation}
where we will refer to $s$ as the {\it scale} of the MHW. In 
Figure~\ref{fig_MHW} we show the particular case of $s=2'$ in a grid of size
$20' \times 20'$.  Although the scale used to build this MHW was only
2 arcmin, the MHW extends up to several arcmin (beyond the antenna
FWHM of the 353 GHz channel).
In a few words, the effect of the MHW is to magnify, with respect to
the background, the signals with scales around the scale of the MHW.
The idea is to look at the number of wavelet coefficients above a
certain level (threshold) at different scales in the map of CMB plus
SZE residual (plus noise) and compare these coefficients with the ones
obtained when only the CMB (plus noise) is considered. We show the
result for the channel at 353 GHZ in 
Figure~\ref{fig_N_Threshold_MHWF}, where we have changed the scale just to
see how the contribution of the SZE residual changes with the scale of
the MHW. In our simulations, we have also included the expected Planck
noise level in this channel.  The instrumental noise is simulated as a
white Gaussian noise with an RMS of $14.4\times10^{-6}$ per resolution element 
in $\Delta T/T$ units.  The result is compared with the mean value and $1\sigma$
error bars obtained from 100 Gaussian realizations of the CMB (plus
instrumental noise).  The mean value of these realizations is
consistent with the expected number of wavelet coefficients (in absolute value) 
above the threshold for a map with $512^2$ MHW coefficients. This expected number 
is just $0.0465\%$ (for a threshold = $3.5\sigma$) of $512^2$ which is
$\approx$ 120.  This number is a constant independent of the scale of
the MHW since the convolution of the CMB maps with the MHW does not
change their Gaussian nature. Since the original map contains $512^2$
pixels, the convolved map will contain $512^2$ MHW coefficients and
the expected number of MHW coefficients (in absolute value) above $3.5\sigma$ for a
Gaussian map will remain constant at all scales, $\approx$ 120.  The
reason why we choose a threshold of $3.5\sigma$ is because we have to
find a compromise between having enough statistics and the
significance of the result.  A lower threshold will produce a larger
number of coefficients above the threshold but these number will be
more dominated by the Gaussian part of the distribution.  A higher
threshold will select the tail of the Gaussian distribution which will
show more clearly the non-Gaussian contribution due to the SZE
residual. However, a high threshold will have a low expected number of
coefficients above the threshold in a Gaussian case. We choose the
threshold at $3.5\sigma$ because for this threshold, both the mean
number of Gaussian coefficients and the significance are high enough
(120 and $3.5\sigma$ respectively).
As it can be seen in figure~\ref{fig_N_Threshold_MHWF}, the CMB plus
SZE residual shows a clear deviation from the Gaussian case 
which is larger when the scale is smaller. The deviation is
maximum at scales of the MHW around 3 arcmin ($4.25 \sigma$ at 2.8 arcmin).  
For the rest of the channels, the non-Gaussian signature of the CMB plus SZE residual is
consistent (within the corresponding error bars) with the Gaussian
case (see the 143 GHz case). This is due to two things. 
First, the amplitude of the SZE residual is smaller in the other 
channels.  Second, and more important, the antenna is
larger and the signal is diluted so the small scales cannot be
magnified with the same efficiency as in the channel at 353 GHz.  
For comparison we show the result when our Gaussianity estimator is 
applied to the 143 GHz channel. In this case, no significant non-Gaussianity 
is observed at small scales. In the rest of the channels, the antenna is 
even larger so the dilution of the residual is bigger. 
The main conclusion is that the 353 GHz channel can contain a significant
non-Gaussian signal at small scales coming for the SZE residual. It is
worth pointing out, that MAP will not measure non-Gaussianity from SZE
residuals mainly because of the lower resolution, $\sim 12$ arcmin at $90$ GHz 
(i.e. the situation will be similar to what we observe in the channels 
below 143 GHz).\\
It is also worth pointing out that such non-Gaussianity studies could be
used to determine which component separation methods are most useful
for Planck. In future works focusing on non-Gaussianity studies, one
could imagine using the 353 GHz channel to remove the non-Gaussianity
due to SZE residuals from the other channels. Another interesting
possibility is to use this non-Gaussianity to remove even more of the
SZE residual. 

\begin{figure}
   \begin{center} \epsfxsize=8.cm
   \begin{minipage}{\epsfxsize}\epsffile{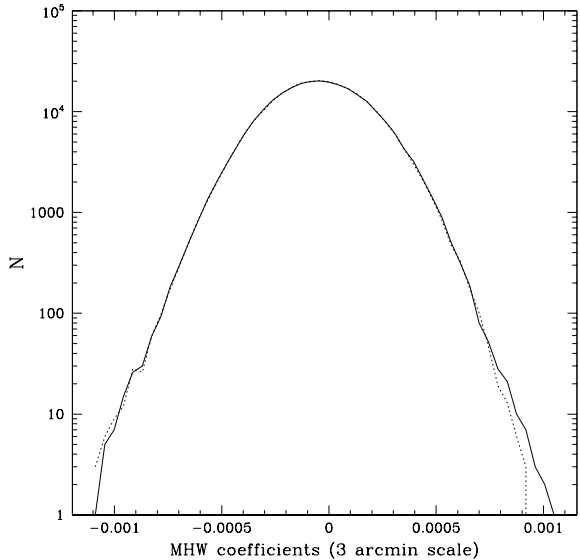}\end{minipage}
   \caption{\label{fig_Histogram_3arcminMHW} Histograms of the 3
   arcmin MHW coefficient maps.  The dotted line is the case when only
   the CMB plus noise is considered (no residual). The solid line shows the excess 
   when the the SZE residual is added to the CMB plus noise map (353 GHz channel).}
   \end{center}
\end{figure}

In Figure~\ref{fig_Histogram_3arcminMHW} we show the histogram of the
two wavelet coefficient maps (CMB plus noise and CMB plus noise plus
SZE residual) for the case when the maps are filtered with a MHW with
a scale of 3 arcmin. This plot shows again the non-Gaussian nature of
the CMB plus SZE residual map at small scales.\\
We have also applied our non-Gaussianity estimator to see if we can detect some 
non-Gaussianity coming from the contribution of the kinematic SZE. We did not 
find any non-Gaussian signatures due to this component. 

\section{Conclusions}\label{section_Conclusions}
In this paper we have studied the systematic errors introduced in the
Planck cluster catalogue and in the CMB map due to the imperfect SZE
subtraction caused by the non-relativistic assumption and to the
intrinsic error in the component separation process.  We have used the
non-parametric method proposed in \cite{Diego2002} to perform the
component separation method. This method is very robust in the sense
that it makes a minimum number of assumptions. The drawback is that it
is not the most precise in the determination of the SZE
component. However, this allow us to put an upper limit on the
systematic errors introduced by the imperfect SZE subtraction since any
more sophisticated method in principle should produce smaller SZE
residuals.

We have seen that the effect of the non-relativistic assumption in the
frequency dependence of the SZE made in the component separation
process is small when compared to the case where the real frequency
dependence is assumed (between 4\% and 8\% relative difference for $T
= 10$ keV clusters), however, the relativistic corrections should be
considered in the simulations of the SZE in order to compute correctly
the selection function and completeness level of the cluster
catalogue.

Concerning the kinematic component of the SZE, relativistic
corrections should be taken into account in order to recover this
component. Otherwise errors as large as $50-100 \%$ could be
introduced in this component in the most relevant channel for its
detection (217 GHz).

In the CMB, the SZE residual in the 353 GHz leaves a non-Gaussian
signature at small scales which could be detected by some Gaussianity
estimators like the MHW. This channel could thus be used to extract
non-Gaussianity signatures from imperfect SZE subtraction from the
other channels.  Our MHW Gaussianity estimator does not show
significant deviations from Gaussianity due to imperfect SZE recovery
in the other channels relevant to the CMB.

The SZE residual does not change significantly the power spectrum of
the CMB in the 353 GHz channel. It is therefore safe to include the
353 GHz channel for the computation of the CMB power spectrum. 
This channel, together with the 217 GHz channel,
are the most important ones which will contribute to the CMB map at
the smallest scales ($\approx 5$ arcmin).

\section*{Acknowledgments}
It is a great pleasure to thank Sergio Pastor and Dmitry Semikoz for
discussions which initiated this work and for useful comments. 
We are grateful to Dmitry Semikoz for providing the SZ data
used for the fitting formulae. JMD and SHH acknowledge support from
Marie Curie Fellowships of the European Community programme {\it
Improving the Human Research Potential and Socio-Economic knowledge}
under contract numbers HPMF-CT-2000-00967 and HPMFCT-2000-00607.



\end{document}